\documentclass[floatfix,groupedaddress]{revtex4}
\usepackage{amsmath,amsthm,amssymb,amsfonts}
\usepackage{amsfonts}
\usepackage{graphicx}
\usepackage{grffile}
\usepackage{float}
\usepackage{array}
\usepackage{multirow}
\usepackage[export]{adjustbox}

\usepackage[a4paper]{geometry}

\begin{document}

\title{The solvent mediated interaction potential between solute particles: Theory and applications }

\author{Mamta Yadav}
\email[]{mamtayadavbb@gmail.com}
\affiliation{Department of Physics Banaras Hindu University, Varanasi-221005, India }
\author{Yashwant Singh}
\email[corresponding author: ]{ singh.yas44@gmail.com}
\affiliation{Department of Physics Banaras Hindu University, Varanasi-221005, India }

\begin{abstract}
In this paper we develop a theory to calculate the solvent mediated interaction potential between solute particles dispersed in a solvent. The potential is a functional of the instantaneous distribution of solute particles and is expressed in terms of the solute-solvent direct pair correlation function and the density-density correlation function of the bulk solvent. The dependence of the direct pair correlation function on multi-point correlations of the solute distribution is simplified with a mean field approximation. A self consistent approach is developed to calculate the effective potential between solute particles, the solute-solvent and the solute-solute correlation functions. The significance of the solvent fluctuations on the range of the effective potential is elucidated. The theory is applied to calculate equilibrium properties of the Asakura-Oosawa (AO) model for several values of solute and solvent densities and for several values of the particles size ratio. The results give a quantitative description of many-body effect on the effective potential and on the pair correlation functions.
\end{abstract}

\maketitle

\section{INTRODUCTION} \label{IN}
Dealing with complex multi-component systems such as colloidal suspensions, macro-molecular solutions etc, where particles size asymmetry is large, it often becomes necessary to simplify the description by using coarse-graining strategies.
In these, one seeks to eliminate degrees of freedom of those components of the system which can be identified as solvent \cite{likos2001effective,lekkerkerker2011depletion,gonzalez1998effective}. A solvent constitutes all those components of the system which act as background and are experimentally either unobservable or of no direct interest. The coarse-graining leads to effective interactions between particles of the surviving components (referred to as solute). If this is done exactly, the effective interaction will account exactly for the effects of the degrees of freedom which have been subsumed. Thus the statistical properties of the solute will be identical in both the coarse-grained and the full system description.

In principle, one can derive the coarse-grained (or effective) Hamiltonian by integrating out all variables belonging to solvent particles from the system partition function \cite{chandler1984excess,dijkstra1999phase}. For a binary mixture, a formal expression for the effective Hamiltonian was derived by Dijkastra et. al \cite{dijkstra1999phase} by expanding the partition function in powers of the Mayer's functions associated with pair potentials between solvent particles and between solute and solvent particles and organising terms of the expansion according to number of solute particles. The resulting Hamiltonian consists of zero-body, one-body, two-body and many-body interactions and have to be determined one-by-one. This practically restricts the Hamiltonian as a sum of (effective) pair potential obtained by considering a single pair of solute particles along with zero and one body terms \cite{dijkstra1999phase,ashton2011depletion}. However, even though the underlying interactions in the full model are pairwise additive, the coarse-graining will lead to effective potential which is many-body in character. Obtaining a full many body effective Hamiltonian remains, even for a simple case of binary mixture of spherical particles with short range pair interactions an open challenge to theory as well as to computer simulations. As far as simulation is concerned, the required computational investment for a highly particle size asymmetric mixture is generally prohibitive because of very slow relaxation of big particles caused by solvent particles \cite{ashton2011depletion}.

In the last few decades a variety of theoretical methods which include perturbation theory \cite{lekkerkerker1993spinodal,mao1995depletion}, integral equation theory (IET) \cite{mendez2000depletion,castaneda2006entropic,gonzalez2005reference}, density functional theory (DFT) \cite{cuesta1999density,schmidt2002density} have been used to find effective potential between solute particles in a mixture. In an approach initiated by Mendez-Alcaraz and Klein \cite{mendez2000depletion} effective interaction between solute particles is accounted for by a contraction of the description in the framework of IET of simple liquids. However, tackling asymmetric mixtures via IET, where one treats all species on equal footing, is notoriously difficult \cite{ashton2011depletion,amokrane2005structure}. In case of DFT the projection of free-energy functional of mixture onto that of a one-component fluid is difficult as at no stage in the calculation have integrals been performed over the solvent degrees of freedom \cite{schmidt2002density}. The effective potential does not appear explicitly within the DFT formulation. However, DFT has been very successful in calculating the effective (depletion) potential between two big hard spheres in a reservoir of small hard spheres \cite{roth2000depletion,boctan2009hard,oettel2009depletion}. This is because in this particular case one requires a DFT for the solvent only as big particles are fixed, so they simply exert an external potential on the small ones and for a one component hard-sphere system accurate free energy functional exist \cite{rosenfeld1989free}.

In this paper we describe a general theory for the solvent induced interaction potential between solute particles. The theory is based on a formalism developed by one of us \cite{singh1987molecular} to find effective interaction between monomers of a polymer chain dissolved in a solvent. Here we extend the theory and apply it to calculate effective potential between colloidal particles suspended in a sea of solvent particles. In Sec.\ref{THEORY}  we start with the partition function of a system consisting of solute and solvent particles and derive expression for the solvent induced interaction between solute particles by integrating out co-ordinates of solvent particles. The resulting expression is expressed in terms of solute-solvent direct pair correlation function and the density-density correlation function of the pure solvent. An essential feature of this derivation is that the solute-solvent correlation function depends on the state of the solute (i.e distribution of solute particles). An integral equation is derived to calculate the solute-solvent correlation functions. In Sec.\ref{Results} we apply the theory to a system described by the Asakura-Oosawa model \cite{asakura1954interaction,asakura1958interaction,vrij1976polymers} in which solute-solute and solute-solvent interactions are hard-sphere like, but the solvent-solvent interaction is zero (perfectly interpenetrating spheres). As this model captures many features of real colloid-polymer mixtures very well, its statistical thermodynamics continues to be the subject of investigation \cite{brader2003mol,binder2014perspective}. The paper ends with a brief discussion given in Sec.\ref{D}.

\section{\bf{THEORY}} \label{THEORY}
\setcounter{equation}{0}
\renewcommand{\theequation}{2.\arabic{equation}}
We consider a binary mixture of particles of species $a$ (to be called solute) dispersed in a fluid of particles of species $b$ (to be called solvent). For the sake of simplicity we assume that particles of both species are spherically symmetric and have only one interaction site. Generalisation to  non-spherical particles and to a multi-component solvent is straightforward.

The potential energy of interactions between particles are taken to be pairwise sum as
\begin{align} 
\begin{split}
	U_{aa} \bigl{[} \vec{R}^{N_a}\bigr{]}&={\sum_{i<j}^{N_a}} u_{aa}(|\vec{R}_{i}-   \vec{R}_{j}|), \\
	U_{bb} \bigl{[} \vec{r}^{N_b}\bigr{]}&={\sum_{i<j}^{N_b}} u_{bb}(|\vec{r}_{i}-\vec{r}_{j}|), \\
	U_{ab} \bigl{[} \vec{R}^{N_a},\vec{r}^{N_b} \bigr{]}&={\sum_{i=1}^{N_a}}{\sum_{j=1}^{N_b}}u_{ab}(|\vec{R}_{i}-\vec{r}_{j}|),
\end{split}
\end{align} 
 where $N{_a}$ and $N{_b}$ are number of particles in the system, $\vec{R_i}$ and $\vec{r_i}$ are position vectors of solute and solvent particles, respectively. The system is contained in a volume $V$ with number densities $\displaystyle{\rho{_a}=\frac{N{_a}}{V}}$ and $\displaystyle{\rho{_b}=\frac{N{_b}}{V}}$. The canonical partition function of the system can be written as
\begin{equation} 
Z(N{_a},N{_b},V,T)={\frac{1}{N{_a}!\
{\Lambda{_a}^{3N{_a}}}}}Tr{_a}\exp{{\biggl[}W\textbf{[}\vec{R}\textbf{]}-\beta\sum_{i<j}u_{aa}(|\vec{R_i}-\vec{R_j}|){\biggr]}},
\end{equation}  
where
\begin{equation}
e^{W\textbf{[}\vec{R}\textbf{]}}={\frac{1}{N{_b}!\ {\Lambda{_b}^{3N{_b}}}}}Tr{_b}\ \exp{{\biggl[}-\beta\sum_{i<j}u_{bb}(|\vec{r_i}-\vec{r_j}|)-\beta\sum_{i=1}^{N_a}\sum_{j=1}^{N_b}u_{ab}(|\vec{R_i}-\vec{r_j}|){\biggr]}}.
\end{equation} 

Here $\Lambda{_a}$ and $\Lambda{_b}$ are thermal wavelengths of species $a$ and $b$, respectively, and $\beta$ is the inverse temperature in units of the Boltzmann constant $k{_B}$. The trace $Tr{_a}$ is short for the volume integral $\int_{V}d\vec{R}^{N_a}$ over the coordinates of particles of species $a$ and similarly for $Tr{_b}$. $W$ is the reduced free energy,
\begin{equation}
W=-A{_b}-{\Delta A\textbf{[}\vec{R}\textbf{]}},
\end{equation} 
where $A{_b}$ is the reduced (in units of $\beta^{-1}$) Helmholtz free energy of pure solvent (i.e. in absence of solute particles) and $\Delta A$ is the reduced excess free energy arising due to interactions between solvent and solute particles (i.e. due to $U_{ab}$). Since the position vectors $\vec{R}$ of solute particles are held fixed when integration over coordinates of solvent particles
 in Eq.(2.3) is performed, $\Delta A$ depends on the constrained position vectors of solute particles. A single particle density operator which defines the constrained spatial configuration of solute particles can be written as 
\begin{equation}
{\hat\rho_{a}(\vec{R})=\sum_{i=1}^{N_a}\delta (\vec{R}-\vec{R}_{i})},
\end{equation} 
 where $\delta$ is the Dirac function.

An alternative though equivalent definition of $\Delta A$ is the solvent contribution to the potential of mean force (in units of $\beta^{-1}$) or simply, free energy surface for $N{_a}$ solute sites in the system. The problem of determining the effective Hamiltonian therefore reduces to finding the solvent induced free energy surface $\Delta A\textbf{[}\vec{R}\textbf{]}$. 

\subsection{Determination of the free energy surface $\Delta A\textbf{[}\vec{R}\textbf{]}$}
In the absence of solute particles, solvent is a homogeneous system with a position independent density $\rho_b$. But due to presence of solute particles in the system the solvent becomes inhomogeneous with position dependent single particle density $\rho_b(\vec{r})$. The change in density $
\delta \rho_{b}(\vec{r})=\rho_{b}(\vec{r})-\rho_{b}$ at position $\vec{r}$ in the solvent can be associated with a potential field  $\phi_{b}(\vec{r})$ (in units of $\beta^{-1}$) defined as
\begin{equation} 
\frac{\delta \Delta A[\vec{R}]}{\delta \rho_{b}(\vec{r})}=\phi_{b}(\vec{r}).
\end{equation} 
The functional derivative of $\Delta A\textbf{[}\vec{R}\textbf{]}$ is taken at constant temperature and volume. Since the field $\phi_{b}$ is produced by solute particles, it is functional of $\hat\rho_{a}(\vec{R})$ and can be expressed in terms of a functional which couples a tagged solvent particle to the solute density field $\hat\rho_{a}(\vec{R})$ as

\begin{equation} 
\begin{split}
\phi_{b}(\vec{r})=-\int d\vec{R}\ c_{ab}(\vec{r},\vec{R})\ \hat\rho_{a}(\vec{R}),
\\
{=-\sum_{i=1}^{N{_a}} c_{ab}(\vec{r},\vec{R}_{i})} \ ,
\end{split} 
\end{equation} 
where function $c_{ab}$ determines the strength of the coupling. Eq.(2.7) can also be expressed as 
\begin{equation}
	\frac{\delta \phi_{b}(\vec{r})}{\delta \hat\rho_{a}(\vec{R})}=-c_{ab}(\vec{r},\vec{R}).
\end{equation}
Though we call the function $c_{ab}(\vec{r},\vec{R})$ solute-solvent direct pair correlation function, it is not, as shown below in section B, the usual Ornstein-Zernike function of a binary mixture.

Similarly the solvent mediated potential field that acts on a solute particle at $\vec{R'}$ is expressed as  
\begin{equation}
	\frac{\delta\Delta A\textbf{[}\vec{R}\textbf{]}}{\delta\hat\rho_{a}(\vec{R'})}=\psi_{a}(\vec{R'};\textbf{[}\rho_{b}(\vec{r})\textbf{]}),
\end{equation}  
Here the functional dependence of $\psi_{a}$ on $\rho_{b}(\vec{r}) $  is shown explicitly by the square bracket. Since $\Delta A$ is zero at zero solute density, the functional integration of Eq.(2.9) gives
\begin{equation}
\Delta A = \int\ d\vec{R'}\ \hat\rho_{a}(\vec{R'}) \psi_{a}(\vec{R'},\textbf{[}\rho_{b}(\vec{r})\textbf{]})=\sum_{i=1}^{N_a}\psi_{a}(\vec{R}_{i},\textbf{[}\rho_{b}(\vec{r})\textbf{]}).
\end{equation} 
The functional Taylor expansion about the average solvent density $\rho_{b}$ gives a series ordered in powers of the change in the average density. Thus,
\begin{equation} 
 \Delta A =\int\  d\vec{R'}\ \hat\rho_{a}(\vec{R'}) \psi_{a}(\vec{R'};\rho_{b})-\int d\vec{R'}\int d\vec{r}\ \hat\rho_{a}(\vec{R'}) c_{ab}(\vec{r},\vec{R'};\rho_{b})\delta \rho_{b}(\vec{r}) +.... ,
 \end{equation} 
where
\begin{equation}
	c_{ab}(\vec{r},\vec{R'};\rho_{b})=-\displaystyle\frac{\delta \psi_{a}(\vec{R'})}{\delta \rho_{b}(\vec{r})}\Biggr|_{\rho_{b}(\vec{r})=\rho_{b}},
\end{equation} 
and $\psi_{a}(\vec{R},\rho_{b})$ is the potential field exerted on a solute particle at position $\vec{R}$ by the homogeneous solvent of density $\rho_{b}$. Note that the solute-solvent direct pair correlation function defined by Eq.(2.8) is functional of $\rho_{b}(\vec{r})$ whereas the one defined by Eq.(2.12) is simply function of $\rho{_b}$. The other point to be noted is that the higher order terms in Eq.(2.11) involve three and higher-body solute-solvent direct correlation functions. The need to consider these higher order terms, however, arise only when inhomogeneity in solvent measured by $\delta \rho(\vec{r})$ becomes large, that is when the response of solvent to the field created by solute particles becomes nonlinear (see Eq.(2.14)).

 The value of $\psi_{a}(\vec{R},\rho_{b})$ is found by functional integration of Eq.(2.12) which gives 
\begin{equation}
\psi_{a}(\vec{R},\rho_{b})=-\rho_{b}\int_{0}^{1}d\lambda\int d\vec{r} \ c_{ab}(\vec{r},\vec{R};\lambda\rho_{b}).
\end{equation} 
$\psi_{a}(\vec{R},\rho_{b})$ is the reversible work done in placing a solute particle at position $\vec{R}$ in otherwise a uniform solvent of density $\rho{_b}$.

Assuming that the solvent responds linearly to the solute produced potential field, we can write 

\begin{equation}
\begin{split}
	\delta\rho_{b}(\vec{r})=\int \ d\vec{r'} \  \frac{\delta\rho_{b}(\vec{r})}{\delta(-\phi_{b}({\vec{r'}}))}	\Biggr|_{\phi_{b}=0}(-\phi_{b}(\vec{r'})),\\ 
\\
=\sum_{i=1}^{N_{a}}\int  d\vec{r'}\ \chi_{bb}(\vec{r},\vec{r'})\ c_{ab}(\vec{r'},\vec{R}_{i}). 
\end {split}
\end{equation} 
Here use has been made of Eq.(2.7) for $\phi_{b}(\vec{r})$.
	$\chi_{bb}(\vec{r},\vec{r'})=\ < \delta \rho_{b}(\vec{r})\delta \rho_{b}(\vec{r'})> $  is the density-density correlation function of the bulk solvent and is expressed as \cite{hansen2006theory}
\begin{equation} 
	\chi_{bb}(\vec{r},\vec{r'})\ =\ \frac{\delta\rho_{b}(\vec{r})}{\delta(-\phi_{b}({\vec{r'}}))}	\Biggr|_{\phi_{b}=0}= \ \rho_{b}\ \delta (\vec{r}-\vec{r'})+\rho_{b}^{2}\ h_{bb}(|\vec{r}-\vec{r'}|),
\end{equation} 
where $h_{bb}(|\vec{r}-\vec{r'}|)$  is the total pair correlation function of the bulk solvent of density $\rho_{b}$. 

Substituting Eq.(2.14) into Eq.(2.11) we get
\begin{equation}
\Delta A(\textbf{[}\hat\rho_{a}\textbf{]},\textbf{[}\rho_{b}\text{]}) = 
\sum_{i=1}^{N_{a}}\psi_{a}(\vec{R}_{i},
\rho_{b})\ +\ \frac{1}{2} \sum_{i,j}^{N_a}v(\vec{R}_{i},\vec{R}_{j}) , 
\end{equation} 
where

\begin{equation}
v(\vec{R}_{i},\vec{R}_{j})=-\int d\vec{r}\int d\vec{r'}\  c_{ab}(\vec{r},\vec{R}_{i})\ \chi_{bb}(\vec{r},\vec{r'})\  c_{ab}(\vec{r'},\vec{R}_{j}).
\end{equation}
	The factor $\displaystyle\frac{1}{2}$ is introduced to avoid counting of a pair twice. The first term of Eq.(2.16) is the reversible work required to insert $N_{a}$ solute molecules in the solvent. The second term represents the solvent mediated potential energy of solute. Despite its outward appearance, the expression for $\Delta A[\vec{R}]$ is not pair decomposable since $c_{ab}(\vec{r},\vec{R})$ is a complicated functional of distribution of solute particles.

When Eq.(2.16) is substituted into Eq.(2.2) one gets
\begin{equation}  
Z(N{_a}N{_b},V,T)= e^{-A_{b}}\ Z_{a}(N{_a},V,T),
 \end{equation} 
   where
\begin{equation}
   Z_{a}(N{_a},V,T)={\frac{1}{N{_a}! {\Lambda{_a}^{3N{_a}}}}}Tr{_a}\ e^{ \sum\limits_{i=1}^{N_{a}}\psi_{a}(\vec{R}_{i},\rho_{b})}\ \exp\biggl{[}-\beta\sum_{i<j}U_{aa}^{eff}(\vec{R}_{i},\vec{R}_{j})\biggr{]}
   \end{equation}
is the coarse-grained partition function of the solute. The effective potential between two solute particles is
   \begin{equation}
U_{aa}^{eff}(\vec{R}_{i},\vec{R}_{j})=u_{aa}(|\vec{R}_{i}-\vec{R}_{j}|)\ +\ k_{B}T \ v(\vec{R}_{i},\vec{R}_{j}) .
\end{equation}
	Since  $ c_{ab}$ that appears in expression of $v(\vec{R}_{i},\vec{R}_{j})$ (see Eq.(2.17))  is functional of $\hat{\rho}_{a}(\vec{R})$ (i.e, depends on the constrained configuration of solute particles), $U_{aa}^{eff}$ depends on the spatial configuration of solute particles. The equilibrium properties of solute can now be calculated from the partition function $Z_{a}(N{_a},V,T)$ given by Eq.(2.19).

\subsection{The solute-solvent correlation functions }
\setcounter{equation}{20}
\renewcommand{\theequation}{2.\arabic{equation}}
To find an expression for $c_{ab}$ we adopt a method suggested by Percus \cite{percus1962approximation}. In particular, we fix a solute particle (denoted as $1$) at the origin and calculate effect of its potential field on the density. The change in the solvent density at position $\vec{r}$ is given as \cite{singh1987molecular} 
\begin{equation}
	\delta \rho_{b}(\vec{r}/u_{ab}(\vec{r}))=\sum_{i}\int d\vec{r'}\ \chi_{bb}(\vec{r},\vec{r'})\ c_{ab}(\vec{r'},\vec{R}_{i})\ \ \hat{\rho}_{a}(\vec{R}_{i}/u_{aa}(\vec{R}_{i})),
 \end{equation} 
	where $u_{ab}(\vec{r})$ is the pair potential between solute particle $1$ fixed at the origin and a solvent particle at position $\vec{r}$ and $u_{aa}(\vec{R}_{i})$ is the pair potential between solute particle $1$ and the $i^{th}$ particle at position $\vec{R}_{i}$. It is known that \cite{singh1987molecular,percus1962approximation}

\begin{equation}
\delta \rho_{b}(\vec{r}/u_{ab}(r))=\rho_{b}\ h_{ab}(\vec{r}),
\end{equation} 
where $h_{ab}(\vec{r})$ is the total solute-solvent pair correlation function, and 

\begin{equation}
\rho_{a}(\vec{R}_{i}/u_{ab}(\vec{R}_{i}))= w_{1i}(\vec{R}_{i})\ .
\end{equation} 
Here $w_{ij}(\vec{R}_{i},\vec{R}_{j})$ denotes at two particles level the constrained distribution of solute particles and can be expressed as
\begin{equation}
w_{1j}(\vec{R}_{j})=\delta_{1j}\delta(\vec{R}_{j})\ +\ g_{aa}(\vec{R}_{j}),
\end{equation} 
where $ g_{aa}(\vec{R})$ gives joint probability of locating two different solute particles at position $\vec{R}$  apart in the constrained distribution.
When above results are substituted in Eq.(2.21) one gets 
\begin{equation}
\begin{split}
	h_{ab}(\vec{r})=c_{ab}(\vec{r})\ + \ \rho_{b}\int d \vec{r'} \ h_{bb}(|\vec{r}-\vec{r'}|) \ c_{ab}(\vec{r'}) \ +\ \sum_{i=1}^{N_{a}} \ c_{ab}(\vec{r},\vec{R}_{i})\ g_{aa}(\vec{R}_{i})\ \\
  +\ \rho_{b}\sum_{i=1}^{N_{a}}  \int d \vec{r'} \ h_{bb}(|\vec{r}-\vec{r'}|) \ c_{ab}(\vec{r'},\vec{R}_{i})\ g_{aa}(\vec{R}_{i}). 
\end {split}
\end{equation} \\
	This expression can also be arrived at from the relation (see Eq(2.16)) 
\begin{equation}
\frac{\delta(-\Delta A)}{\delta c_{ab}(\vec{r},\vec{R})}=\rho_{b}\ \hat\rho_{a}(\vec{R})\ g_{ab}(\vec{r},\vec{R}) ,
\end{equation}
	where, $g_{ab}(\vec{r},\vec{R})=1\ +\ h_{ab}(\vec{r},\vec{R})$\ .

Equation(2.25) combined with Eq.(2.17) suggests that for each fixed solute configuration one should determine $c_{ab}$ by solving Eq.(2.25) and then use it in Eq.(2.17) to find $v(R)$. This will lead to different values of $v(R)$ for each fixed solute configuration. To find the most probable value of $v(R)$ and therefore the most probable solute configuration one has to solve the partition function of Eq.(2.19) in which each case is weighted by the Boltzmann factor.
The complicated nature of this approach can, however, be simplified with the aid of a mean field approximation. In particular, we assume that the primary contributions to the partition function (Eq.(2.19)) come from those constrained solute particle configurations with pair distributions  $g_{aa}(\vec{R})$ which are close to the averaged pair distribution (i.e, corresponds to the saddle point solution). This allows us to replace $g_{aa}(\vec{R}_{i})$ in Eq.(2.25) by the equilibrium pair distribution function  $g_{aa}(R,\rho_{a})$ where $\rho_{a}$ is the averaged density of the solute. The link between $g_{aa}$ , $c_{ab}$ and $v$ is retained by using a self consistent approach (described below). It may be noted that this approximation does not in any obvious way neglect fluctuations in the solvent which are likely to play central role in determining the strength and the range of the solvent induced interactions \cite{maciolek2018collective}.

Since $g_{aa}(R,\rho_{a})$ depends on distance $R$, not on a particular position vector, one can replace summation in Eq.(2.25) by integration. Thus,
\begin{equation}
\begin{split}
	h_{ab}(r)=c_{ab}(r)\ +\  \rho_{b}\int d \vec{r'} \ h_{bb}(|\vec{r}-\vec{r'}|) \ c_{ab}(r')\
	+\ \rho_{a} \int d\vec{R} \ c_{ab}(|\vec{r}-\vec{R}|)\ g_{aa}(R) \\
	+\ \rho_{a}\rho_{b} \int d \vec{R} \int d\vec{r'} \ h_{bb}(|\vec{r}-\vec{r'}|) \ c_{ab}(|\vec{r'}-\vec{R}|)\ g_{aa}(R).
\end {split}
\end {equation}\\
 The correlation function $h_{bb}(r)$ that appears in above equation is the total pair correlation function of pure solvent and is found from the integral equation theory \cite{hansen2006theory}. The pair distribution function $g_{aa}(R)$ has to be calculated from the effective interaction $U_{aa}^{eff}(R)$ using either an integral equation theory or by computer simulation. From known $h_{bb}(r)$ and $g_{aa}(R)$, Eq.(2.27) is solved for $c_{ab}$ and $h_{ab}$ using a suitable closure relation. Note that to calculate $ g_{aa}$ one needs $U_{aa}^{eff}$ which in turn is calculated from Eq.(2.17) and Eq.(2.20) which involve $c_{ab}$. Therefore, one has to adopt an iterative method to find self-consistent values of $c_{ab}$,\ $h_{ab}$,\ $g_{aa}$,\ $U_{aa}^{eff}$.
	The correlation function $g_{ab}(\vec{r})=1+h_{ab}(\vec{r})$ gives distribution of solvent particles around a solute particle. The solvent density at position $r$ from a solute particle is $\rho_{b}(\vec{r})=\rho_{b} \ g_{ab}(\vec{r})$.

\section{\bf{Results for the Asakura-Oosawa model}} \label{Results}
\setcounter{equation}{0}
\renewcommand{\theequation}{3.\arabic{equation}}
While the theory developed here will be used to investigate in detail the structural and thermodynamic properties of some model systems in our next paper, here we report results for the solvent mediated potential between solute particles  and other related quantities for the Asakura-Oosawa (AO) model. This simple model, originating from the work of Asakura and Oosawa \cite{asakura1954interaction,asakura1958interaction} and Virj \cite{vrij1976polymers}, describes colloidal hard-spheres in a solvent of noninteracting point particles modeling ideal polymers. The solvent particles overlap with zero energy, irrespective of their distance $r$,
\begin{align}
   u_{bb}(r)=0,
\end {align}
while both the overlap between colloidals (species $a$ particles) and colloid and polymer (species $b$ particles) is forbidden,  

\begin{align}	
 u_{aa}(R\leq\sigma_{a})=\infty,\ \ \ \ \ \ \ \  u_{aa}(R>\sigma_{a})=0
\end{align}
and
$$u_{ab}\bigl{(}|R-r|\leq \frac{1}{2}(\sigma_{a}+\sigma_{b})\bigr{)}= \infty , \ \ \ \ \ \ \ \  u_{ab}\bigl{(}|R-r| > \frac{1}{2}(\sigma_{a}+\sigma_{b})\bigr{)}=0 $$
Here $\sigma_{a}$ and $\sigma_{b}$ are, respectively, diameters of solute (colloidal) and solvent particles.

This is an athermal model in which only packing constraints matter, there are no energy parameter whatsoever; all the phase behaviour that result is purely due to entropy. The solvent density is found to play a role like inverse temperature, and the size ratio of solvent versus colloid diameters\ ($\frac{\sigma_{b}}{\sigma_{a}}=q$) acts a as control parameter to modify the phase diagram \cite{dijkstra2006effect,vink2004grand}. The solvent induces an effective attraction (depletion potential) among colloid particles and for sufficiently large size ratio $q$, a liquid-liquid phase separation in a solvent rich phase and a colloid rich phase occurs \cite{binder2014perspective,dijkstra2006effect,vink2004grand}. Of particular interest is many-body effect on the effective attraction. The model has been investigated using theoretical methods such as free volume approximation \cite{lekkerkerker1992phase}  and DFT \cite{schmidt2002density,brader2003mol} and by computer simulation \cite{dijkstra2006effect,vink2004grand}. The usefulness of this model as a generic colloid model and workhorse to explore bulk and inter-facial phenomena in soft matter has been emphasized in a recent review by Binder et. al \cite{binder2014perspective}. Here our aim is limited to calculating many-body effect on the solvent induced interaction between solute particles.\\ 
Since the solvent species $b$ behaves as an ideal gas, $h_{bb}(r)=0$ at all density $\rho_{b}$. When we substitute this in equations Eq.(2.17) and Eq.(2.27) we find,
 \begin{equation}
  v(R) = -\rho_{b} \int \ d\vec{r}\  c_{ab}(r)\  c_{ab}(|\vec{R}-\vec{r}|)
\end{equation}
and 
\begin{align}
h_{ab}(r)=c_{ab}(r)\ +\  \rho_{a}\int \ d \vec{R}\ c_{ab}(|\vec{r}-\vec{R}|) \ g_{aa}(R)\ .
\end{align}
In the limit $\rho_{a}\rightarrow 0$, \ \  $h_{ab}(r)=c_{ab}(r)=e^{-\beta u_{ab}(r)}-1$, is
the Mayer function representing solute-solvent interaction. When we substitute this into Eq.(2.13) and Eq.(3.3) we get following well known results \cite{binder2014perspective,hansen2006theory}.
\begin {equation} 
\psi_{a}= \eta_{b} \Biggl( \frac{ 1+q }{ q } \Biggr)^{3},
\end {equation}
and 
\begin {equation}
\begin{split} 
 v(R)\ &= -\eta_{b}\ \Biggl( \frac{ 1+q }{ q } \Biggr)^{3} \Biggl[1-\frac{3R}{2\sigma_{a}(1+q)}+\frac{r^{3}}{2\sigma^{3}_{a}(1+q)^{3}}\Biggr] \quad  \quad \text{for}  \ \  \sigma_{a}\leq R\leq\sigma_{a}(1+q),
 \\ \\
&=0   \qquad\qquad\qquad\qquad\qquad\qquad \qquad\qquad\qquad\qquad \ \ \ \quad \text{for} \ \  R>\ \ \sigma_{a}(1+q),
 \end {split}
\end {equation}
where \ 
	$ q=\ \displaystyle\frac{\sigma_{b}}{\sigma_{a}}$\  and \ $\eta_{b}=\ \displaystyle\frac{\pi}{6}\rho_{b}\  \sigma_{b}^{3} $ and $\psi_{a}$ and $v$ are in units of $\beta^{-1}$. 

This is an exact expression for depletion potential $v(R)$ for $q < q_{c}=2\sqrt{3}-1\simeq 0.1547$. However, for $q > q_{c}$ it would be a crude approximation as many-body effect becomes nonzero \cite{gast1983polymer}. This can be made plausible by geometric argument, since the number of non overlapping colloidal spheres that can simultaneously overlap with a solvent particle increases when $q$ increases. Using a Monte Carlo scheme, Dijkastra et. al. \cite{dijkstra2006effect} showed that as value of $q$ increases above $q_{c}$, the number of depletion layer that can simultaneously overlap increases. Recently Ashton and Wilding \cite{ashton2014quantifying,ashton2014three} developed a simulation technique based on an approach for determining virial coefficients from the measured volume-dependent asympotate of a certain structural function to calculate difference between the third virial coefficient of the full system described by Eqs.(3.1) and (3.2) and that of a system described by the pair potential $\beta u^{eff}(R)= \beta u_{hs}(R)+v(R)$ where $v(R)$ is given by Eq.(3.6). Santos et.al \cite{santos2015effective}, on the other hand, used a mapping between the first few virial coefficients of the binary nonadditive hard sphere mixture representative of the AO model (Eqs.(3.1) and (3.2)) and those found from the pair potential to access many-body effect. While one can infer from their results that the many-body (limited to three and four bodies) effect makes $v(R)$ less attractive, one cannot determine the change that has taken place in value of $v(R)$ due to change in the density of solute particles and in the particle size ratio at a given solvent density.

We use Eqs.(3.3) and (3.4) to calculate values of $v(R)$ for different values of $q$, $\eta_{b}$ and $\eta_{a}$. Eq.(3.4) is solved using relations,
\begin{subequations}
\begin{equation}
h_{ab} = -1\quad \text{for} \quad r \leq d=\frac{1}{2} \sigma_{a}(1+q) ,
\end{equation}
and
\begin{equation}
c_{ab}=0 \qquad\qquad   \text{for} \quad r>d .
\end{equation}
\end{subequations}
For a system of hard spheres, as is well known \cite{hansen2006theory}, Eq.(3.7a) is exact and Eq.(3.7b) is the Percus-Yevick (PY) approximation. We used a functional defined from Eq.(3.4) as
\begin{equation}
\begin{split}
 I[c_{ab}(r)] &= \int \ d\vec{r}\ (1+h_{ab}(r))\ c_{ab}(r)\\
 &=\int \ d\vec{r}\ (1+c_{ab}(r)) \ c_{ab}(r)\  +\rho_{a} \int \ d\vec{R}  \int  \ d\vec{r}  \ c_{ab}(r) \ c_{ab}(|\vec{r}-\vec{R}|)\ g_{aa}(R),
 \end{split}
\end{equation}\\
and variational relations 

\begin{subequations}
\begin{equation}
0\ =\ \frac{\delta I[c_{ab}(r)]}{\delta c_{ab}(r)}\qquad,\qquad \text{for} \ r\leq d \qquad ,     
\end{equation}

\begin{equation}
c_{ab}(r)\ =\ 0\qquad\qquad,\qquad \text{for}\ r\geq d \qquad ,
\end{equation}
\end{subequations}\\
to find values of $c_{ab}(r)$. We solve Eq.(3.9) numerically by expressing $c_{ab}$ for $r<d$ in a series of basis functions,
\begin{equation}
c_{ab}(r)\ =\ \sum^{n}_{l=0}\ c_{l}\ (d-r)^{l}\ \Theta(d-r)\ ,
\end{equation}
where $\Theta(x)$ is the unit step function which is zero for $x<0$. The functional $I[c_{ab(r)}]$ now becomes function of coefficients $c_{1},c_{2},...c_{n}$ and Eq.(3.9a) reduces to coupled linear equations for these coefficients which is solved numerically.

The only approximation in this numerical solution arises from truncation of the series at a finite $n$. The accuracy of this approximation is checked by computing $h_{ab}$. An exact solution would yield $h_{ab}(r)=-1$ for all $r<d$. This condition led us to choose $n=3$ with error less than $0.01$. In performing calculation we need an expression for $g_{aa}(R)$ which can be found for given $U^{eff}(R)$ using any liquid state theory. Here we used the PY version. This version is very simple and provides an accurate estimate of $g_{aa}(R)$ for the density range of interests in these calculations \cite{hansen2006theory}.

\begin{figure}[H]
	\includegraphics[width=0.5\linewidth ,angle=0,center]{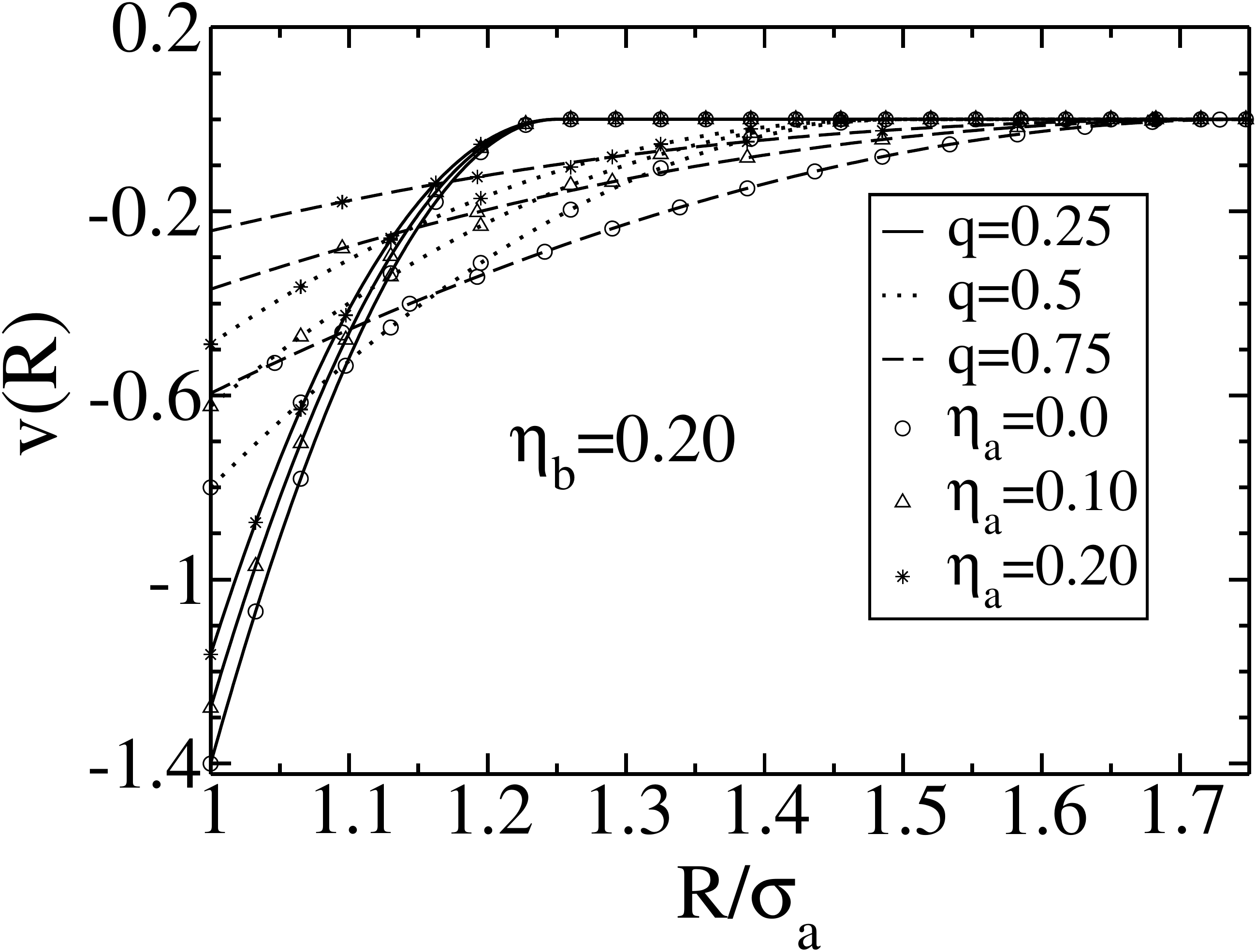}
\caption{The solvent mediated potential $v(R)$ (in terms of $\beta^{-1}$) between a pair of solute (colloid) particles separated by distance $R$ (measured in units of colloid particle diameter $\sigma_{a}$) at solvent packing fraction $\eta_{b}=0.2$ and particle size ratio $q=0.25, 0.5$ and  $0.75$. For each value of $q$, values of $v(R)$ are plotted for three values of solute packing fraction $\eta_{a}=0.0, 0.1$ and $0.2$ with line symbols shown in inset of the figure.}
\label{fig:figure1}
\end{figure}

We plot results found for $v(R)$, $ g_{aa}(R)$ and $ h_{ab}(r)$ for $q = 0.25, 0.50$ and $0.75 $ in Figs.1-9. In Figs.1 and 2 values of $v(R)$ are plotted as a function of separation $R$ (measured in units of colloid particle diameter $\sigma_{a}$ ) for $\eta_{b}=0.2$ and $0.3$ respectively. In each case, results plotted are for $\eta_{a} = 0.0, 0.1$ and $0.2$. In all the cases we find that $v(R)$ remains attractive and its value decreases monotonically on increasing particles separation and becomes zero at $\frac{R}{\sigma_{a}}=(1+q)$. The difference $\Delta v(R,\eta_{a},q)=v(R,\eta_{a},q)-v(R,\eta_{a}=0,q)$ which measures the many-body effect on the potential  $v(R)$ increases on increasing the value of $\eta_{a}$ at fixed values of $q$ and $\eta_{b}$. This effect is also found to depend on values of $q$; as $q$ increases the effect increases. In table $1$ we list values of $\Delta v(\frac{R}{\sigma_{a}}=1,\eta_{a},q)$ found for $\eta_{b}=0.2$ and $0.3$ and for $\eta_{a}=0.1$ and $0.2$ at $q=0.25, 0.5$ and $0.75$. From the results given in the figures and in the table it is obvious that the many-body effect which weakens the effective attraction between colloidal particle becomes important as the density of colloidal particles increases and as the particle size ratio $q$ increases. This increase is due to increase in the number of depletion layers that simultaneously overlap \cite{dijkstra2006effect}. In a recent simulation study \cite{kobayashi2019correction} of a system containing only three colloidal particles and particle size ratio $q=0.4$ it was found that the effect is maximum when overlap of all the three particles with a solvent particle is maximum and decreases when overlap decreases.

\begin{figure}[H]
	\includegraphics[width=0.5\textwidth, angle=0,center]{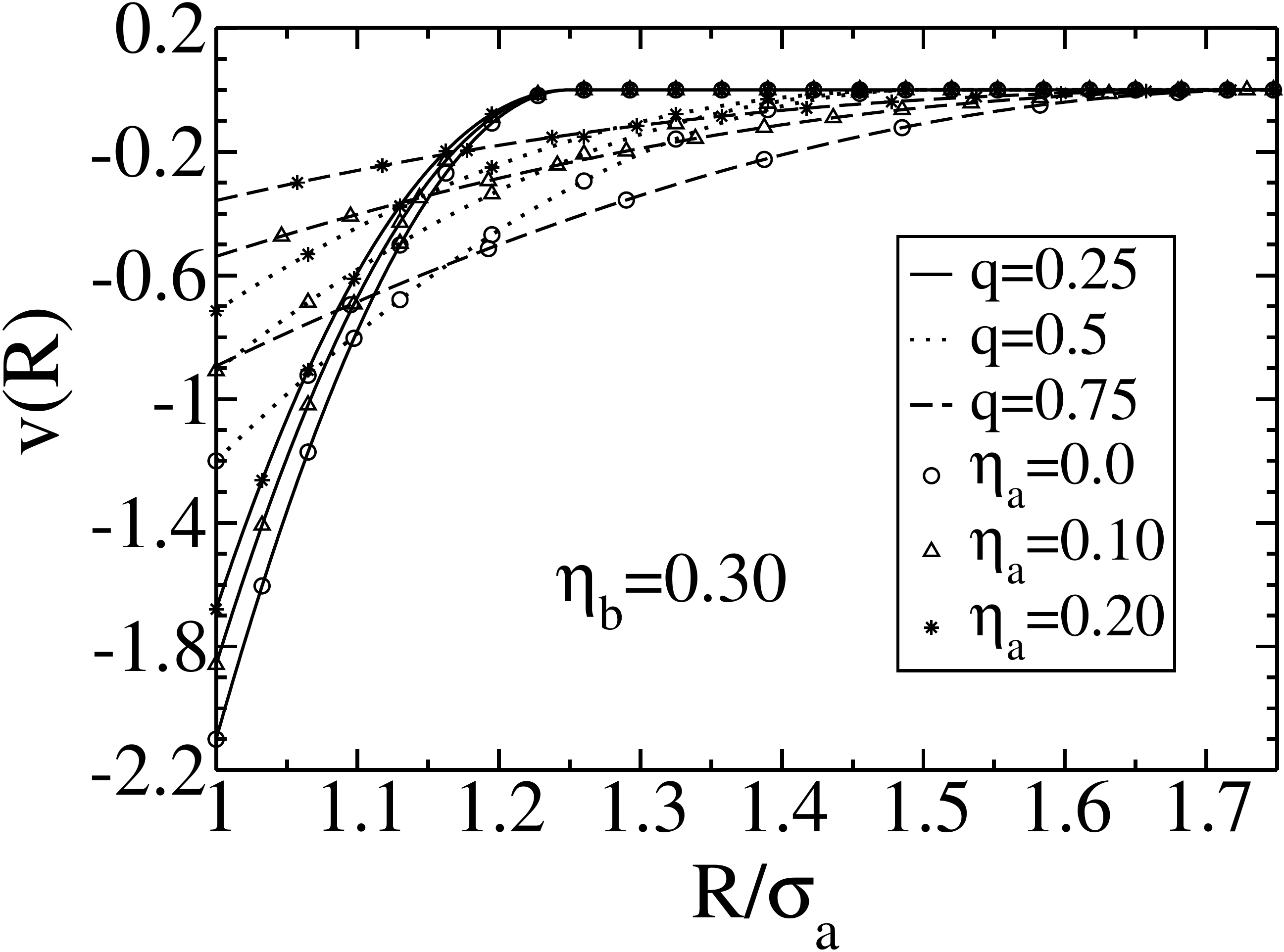}
\caption{As in Fig.1 but now the solvent packing fraction $\eta_{b}$ is fixed at $0.3$.}
\label{fig:figure2}
\end{figure}

\begin{figure}[H]
\includegraphics[width=0.5\textwidth,center]{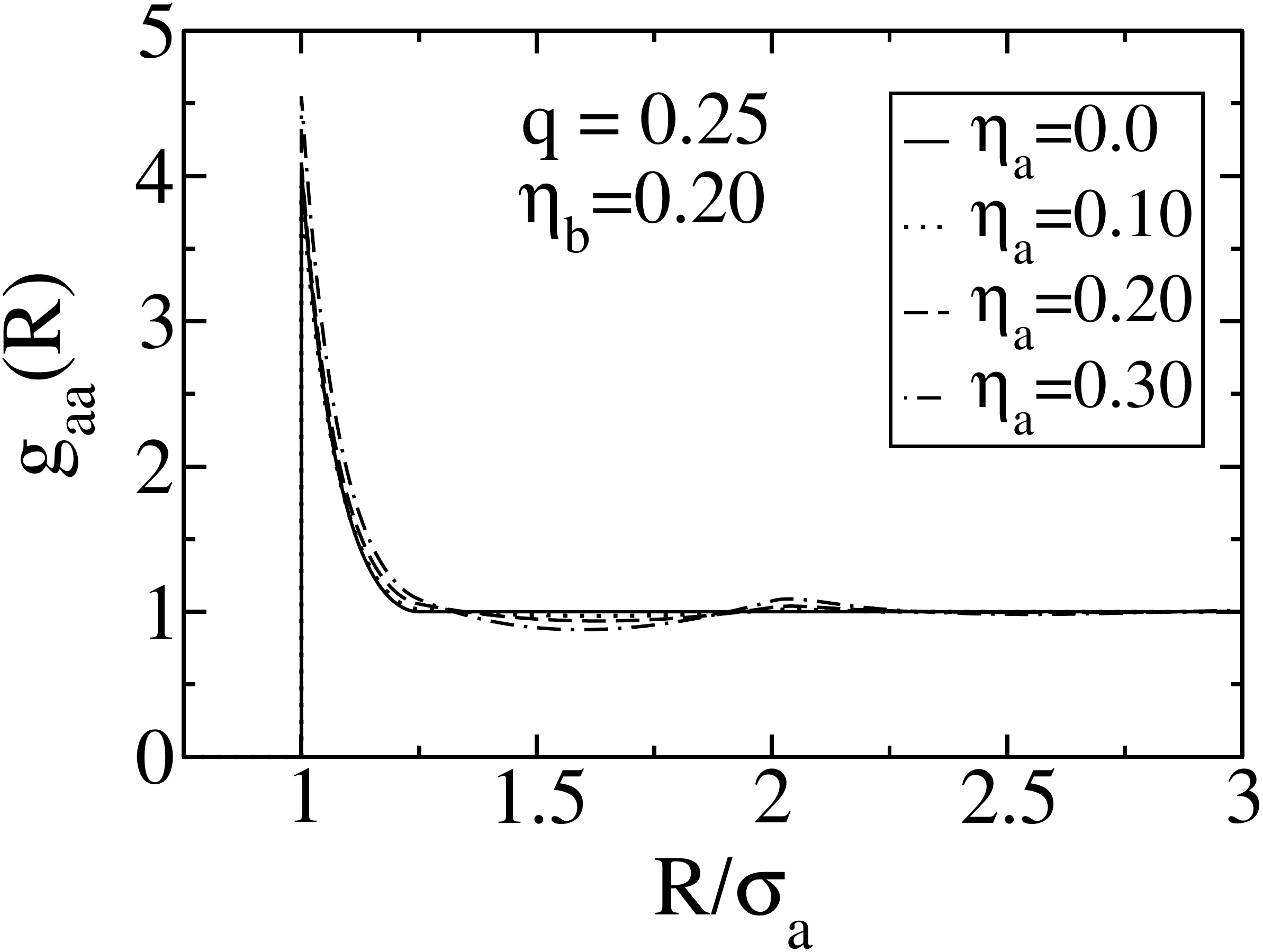}
\caption{Pair distribution function $g_{aa}(R)$ of colloidal particles vs $R$ (measured in units of $\sigma_{a}$) for $q=0.25$ and $\eta_{b}=0.20$ at several values of $\eta_{a}$ shown in the inset.}
\label{fig:figure3}
\end{figure}

\begin{figure}[H]
\includegraphics[width=0.5\textwidth, angle=0,center]{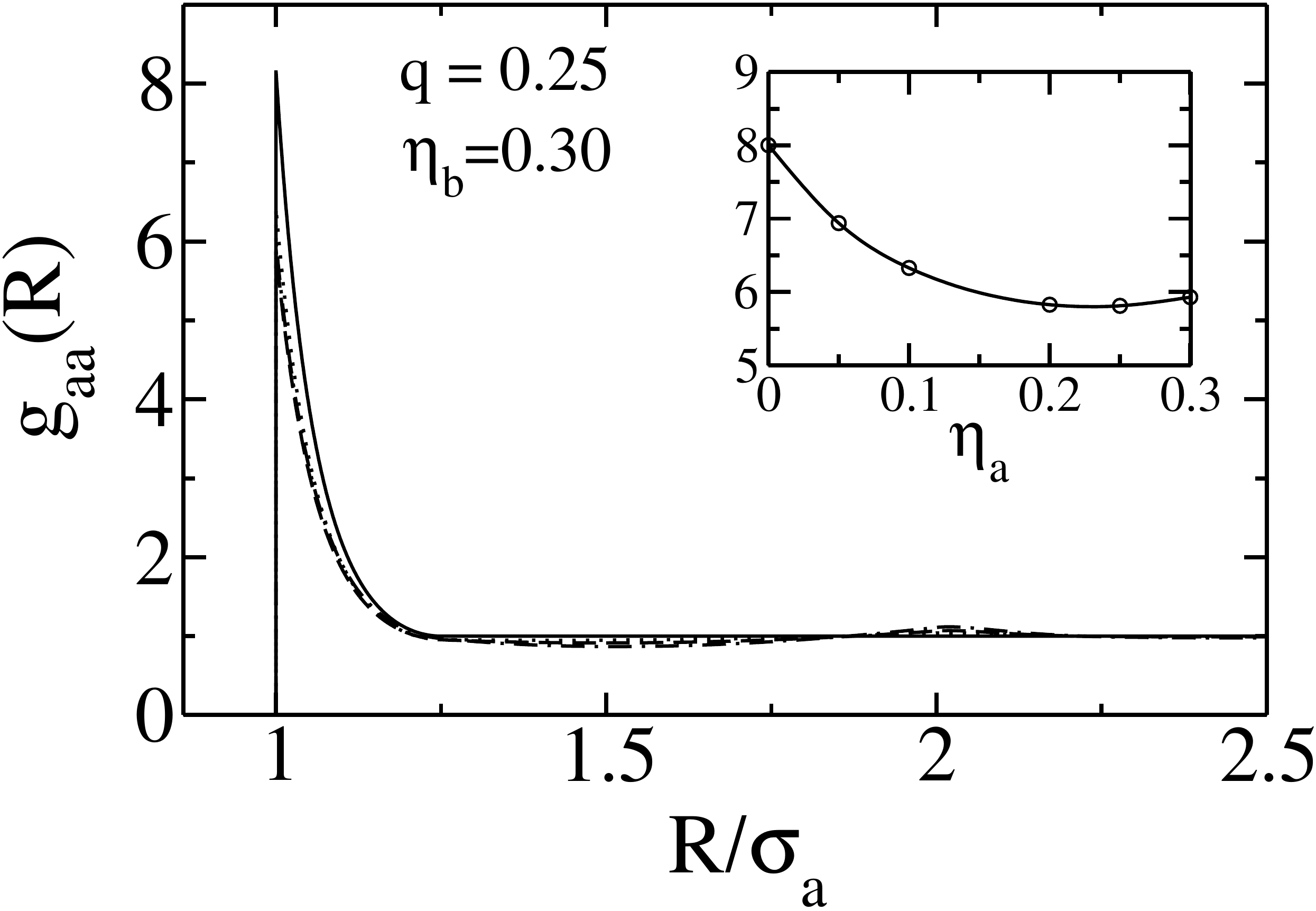}
	\caption{As in Fig.3 but now the solvent packing fraction $\eta_{b}$ is fixed at $0.3$. In the inset the dependence of contact value of $g(R/\sigma_{a}=1)$ is shown; open circles show values found from calculations and the line is drawn to help the eye.}
\label{fig:figure4}
\end{figure}

\begin{figure}[H]
\includegraphics[width=0.5\textwidth,center]{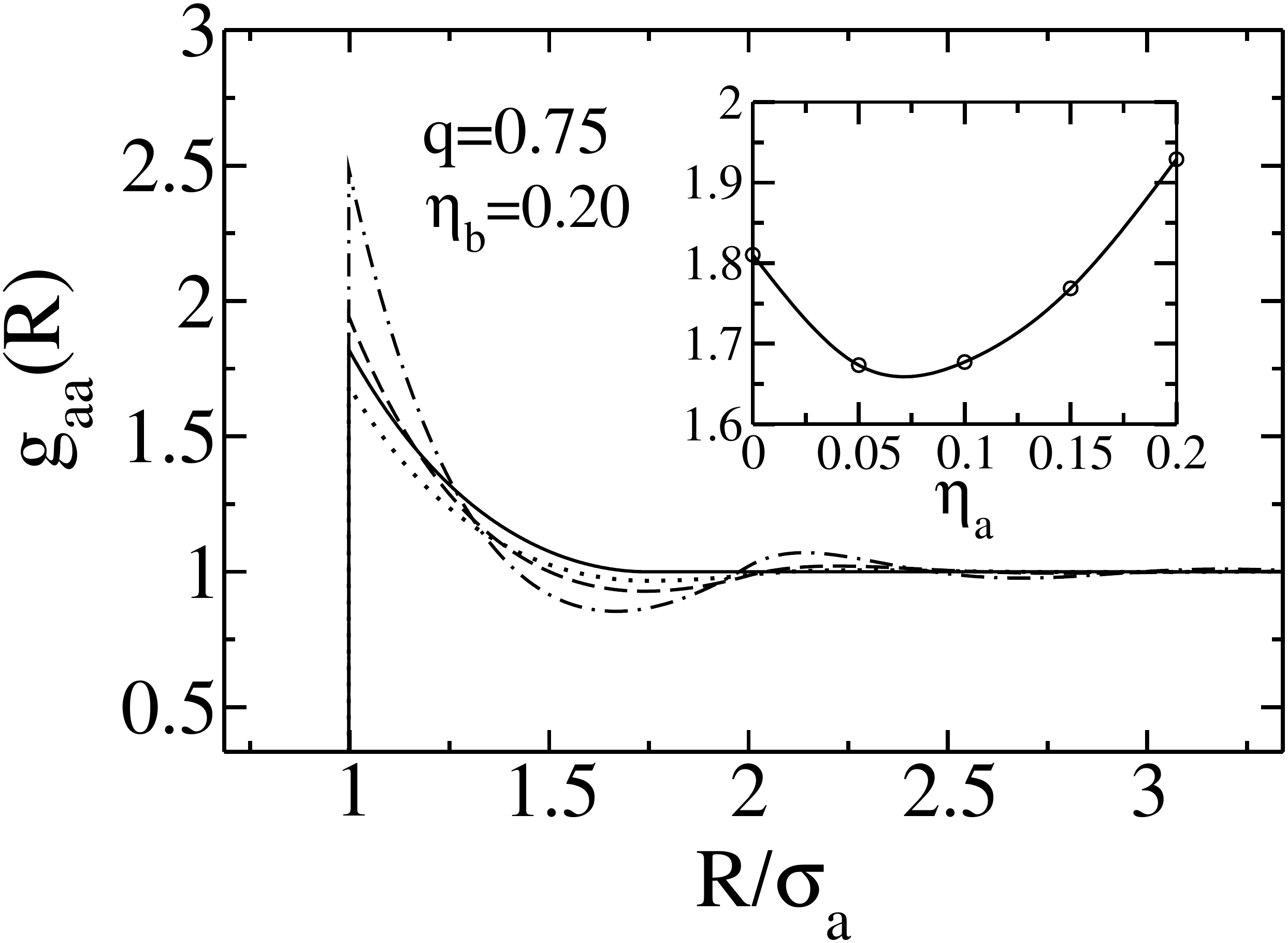}
	\caption{Pair distribution function $g_{aa}(R)$ of colloidal particles vs $R$ for $q=0.75$ and $\eta_{b}=0.20$ at several values of $\eta_{a}$ with lines and symbols same as in Fig.3. In the inset the variation of contact values of, $g_{aa}(R)$ with $\eta_{a}$ are shown. Non-monotonic behaviour of $g(R/\sigma_{a}=1)$ which was absent in the case of $q=0.25$ can be noted.}
\label{fig:figure5}
\end{figure}

\begin{figure}[H]
\includegraphics[width=0.5\textwidth, angle=0,center]{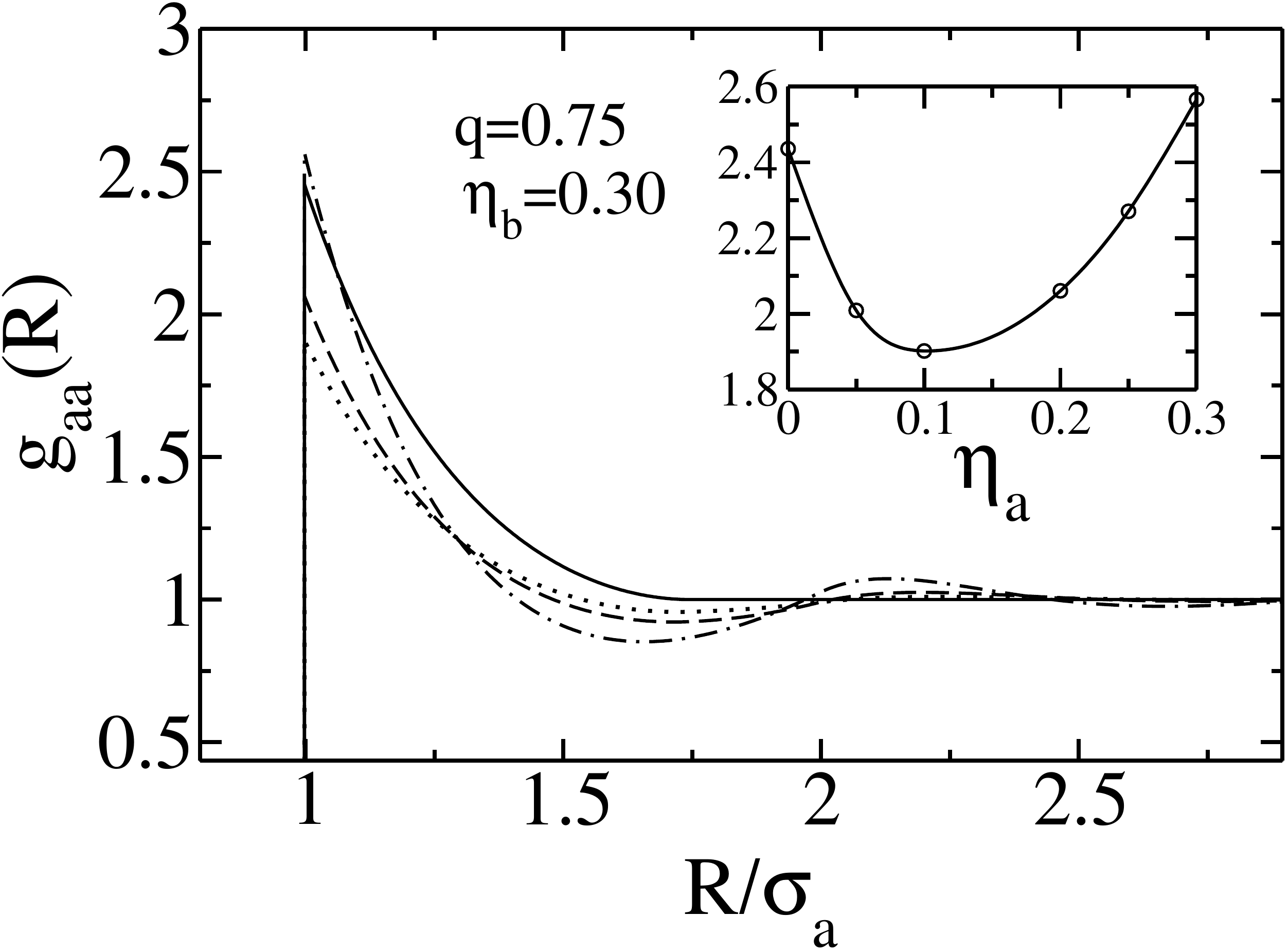}
\caption{As in Fig.5 but now $\eta_{b}=0.3$.}
\label{fig:figure6}
\end{figure}

Values of $g_{aa}(R) $ plotted in Figs.3 and 4 for $q=0.25$ and in Fig.5 and 6 for $ q=0.75$ reveal how pair distribution function of colloid particles depends on value of $\eta_{b},\eta_{a}$ and $q$. Because of the large contact  attraction,the height and width of the first peak of $ g_{aa}(R)$ for $q=0.25 $ is substantially higher and sharper compared to those for $q=0.75$ for the same value of $\eta_{b}$ and $\eta_{a}$. For $q=0.25$ and $\eta_{b}=0.20$ as shown in Fig.3, values of $ g_{aa}(R)$ changes very little as $\eta_{a}$ is increased. However, for $\eta_{b}=0.3$ we find the height of the first peak of $g_{aa}(R)$ decreases as $\eta_{a}$ is changed from zero to $0.3$. For the case of $q=0.75 $ (also for $q=0.5$) values of $g_{aa}(R)$ get substantially modified when $\eta_{a}$ is increased. In particular, we note that the contact value of $g_{aa}(R)$ starts decreasing as the value of $\eta_{a}$ is increased and reaches to its minimum value at a value of $\eta_{a}$ (see inset in Figs.5 and 6) which depends on $\eta_{b}$ and $q$. On further increasing of $\eta_{a}$ the value of $g_{aa}(R/\sigma_{a}=1)$ starts increasing and crosses the value found for $\eta_{a}=0$ at $\eta_{a}\simeq \eta_{b}$.
Should this intriguing behaviour of the pair distribution function of solute particles be linked to the gas-liquid phase separation found by Vink and Horbach \cite{vink2004grand} in a grand canonical Monte Carlo simulation or not, needs further investigation.

\begin{figure}[H]
\includegraphics[width=0.5\textwidth,center]{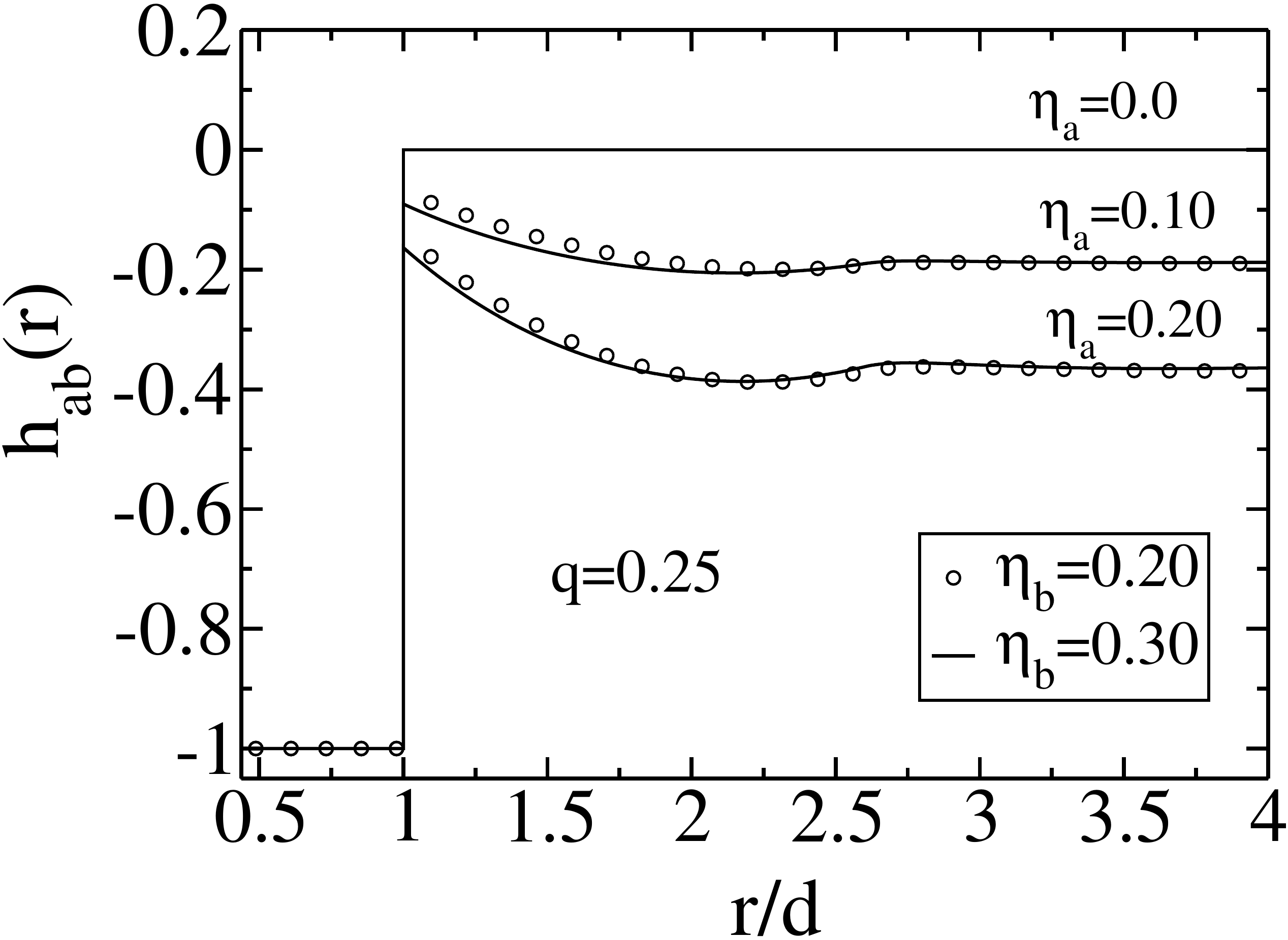}
	\caption{The solute-solvent total pair correlation function $h_{ab}(r)$ vs $r$ (measured in units of $d=\frac{1}{2}\sigma_{a}(1+q)$ ) at $\eta_{b}=0.2$ and $0.3$ for $q=0.25$ and $\eta_{a}=0.0,0.1$ and $0.2$. Values of $h_{ab}$ for $\eta_{b}=0.2$ (circle) and $\eta_{b}=0.30$ (solid line) almost overlap showing that $h_{ab}$ does not depend on $\eta_{b}$. However, $h_{ab}$ is seen to be sensitive to value of $\eta_{a}$.}
\label{fig:figure7}
\end{figure}

\begin{figure}[H]
	\includegraphics[width=0.5\textwidth, angle=0,center]{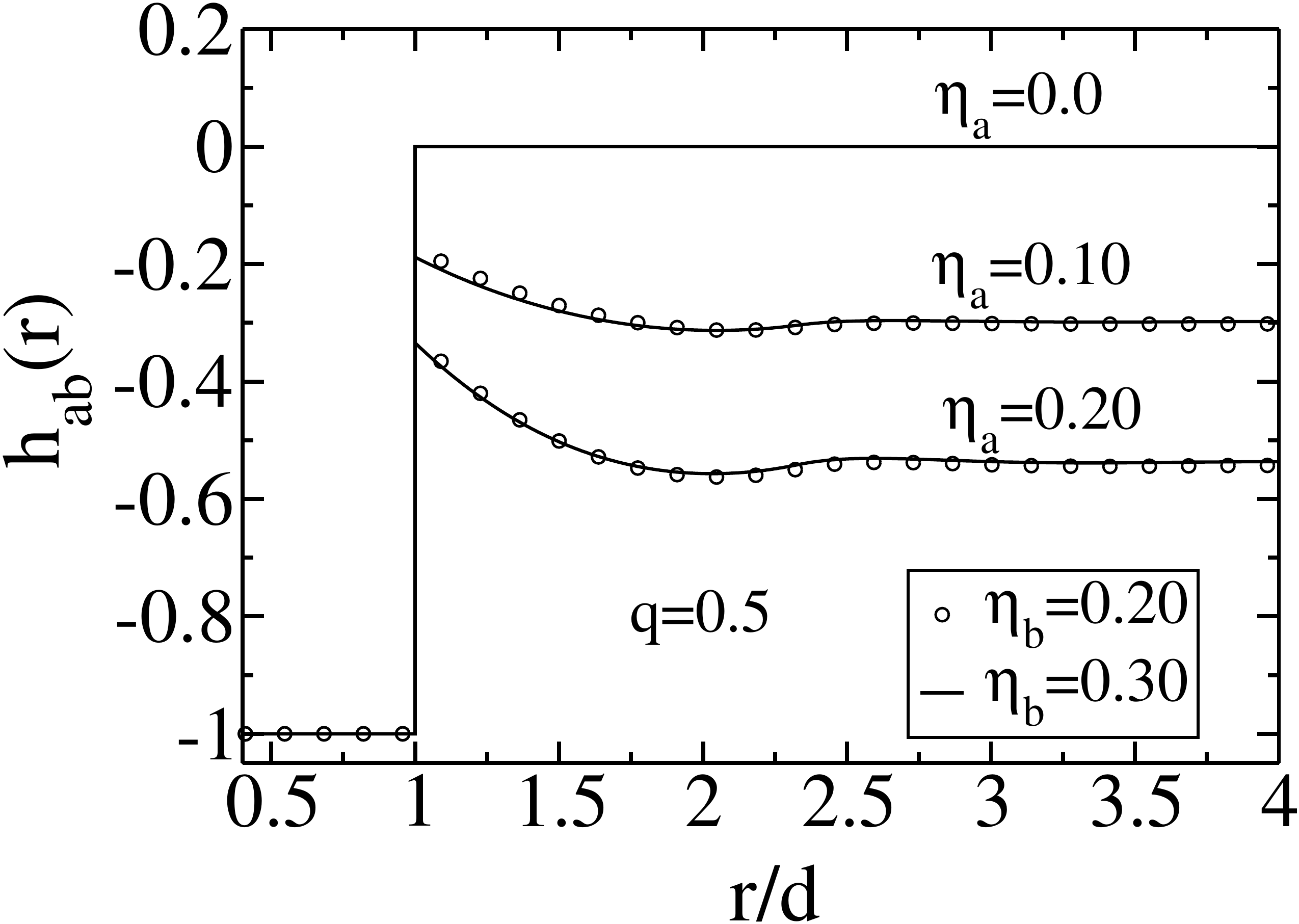}
\caption{As in Fig.7 but for $q=0.5$.}
\label{fig:figur8}
\end{figure}

\begin{figure}[H]
\includegraphics[width=0.5\textwidth, angle=0,center]{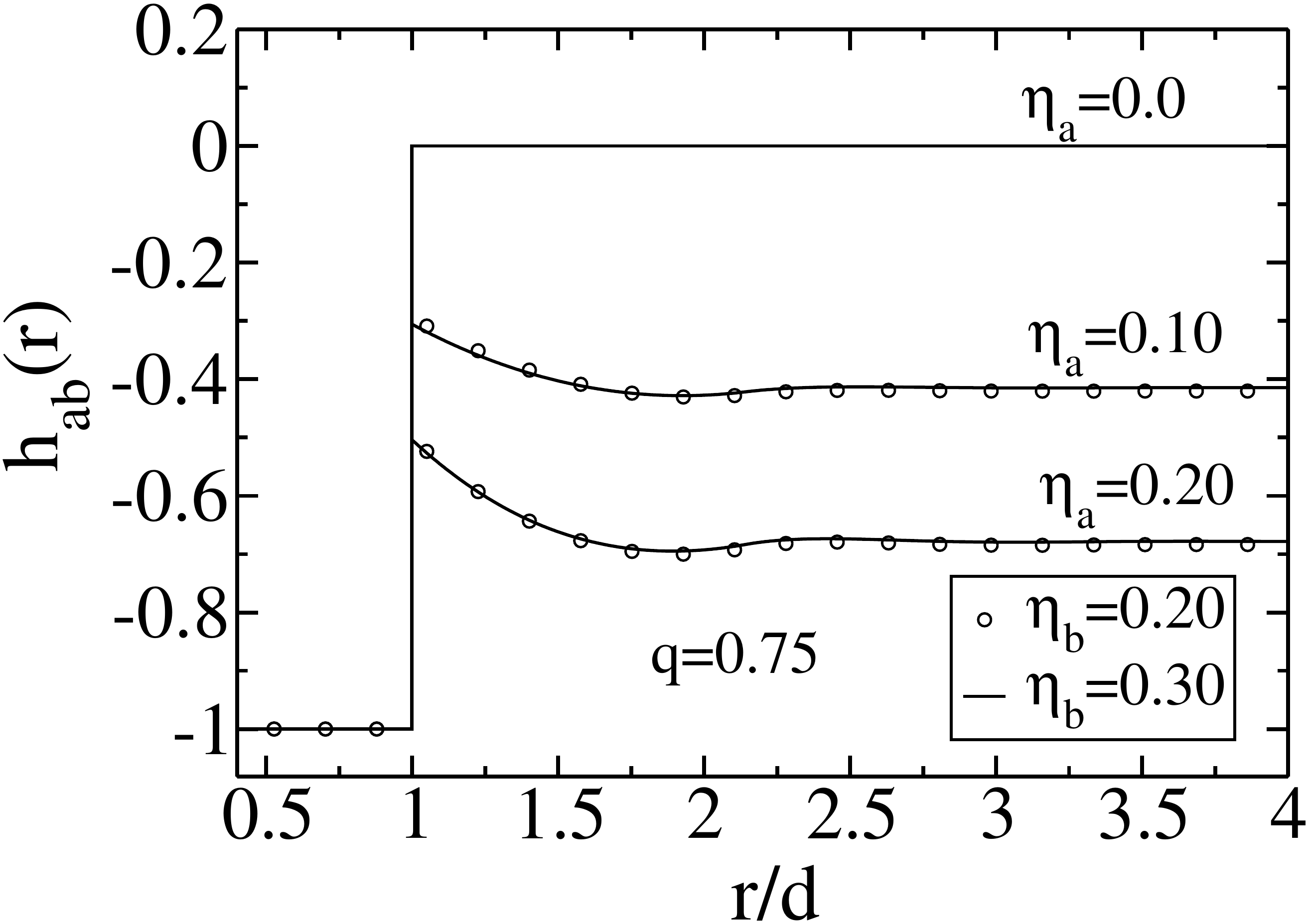}
\caption{As in Fig.7 but for $q=0.75$.}
\label{fig:figur9}
\end{figure}

Values of $h_{ab}(r)$ for $q=0.25,0.5$ and $0.75$ are plotted respectively, in Figs.7-9 as a function of $r$ which measures the separation of a solvent particle from that of the solute. It is found that while value of $h_{ab}(r)$  depends on both, $q$ and $\eta_{a}$, it is nearly independent of $\eta_{b}$; only a small effect is seen close to $r\sim d$. 
The other feature which is perhaps more striking is its value which remains negative and position independent (except near $r\simeq d$). This is due to lack of repulsive interaction between solute particles. It seems that the solvent responds to the increasing number of solute (colloidal) particles by overlapping its particles on each other as no energy cost is involved in doing it. However, as the overlapping reduces entropy, it gets balanced by the entropic force. 

\section{\bf{Discussions}} \label{D}
In the theory described in this paper, the potential field felt by solute particles due to the solvent and the potential field felt by solvent particles due to the solute are expressed in terms of the solute-solvent direct pair correlation function $c_{ab}$. A density functional formalism and the linear response approximation are used to integrate out solvent coordinates from the system partition function. The resulting equations involve $c_{ab}$ which depends on the constrained distribution of solute particles. This complicated functional dependence of $c_{ab}$ on the multi-point solute distributions is simplified with a mean field approximation in which it is assumed that the primary contributions to the effective potential $v$ comes from those distributions which are close to the averaged (most probable) distributions. This reduces the functional dependence of $c_{ab}$ on the averaged two-particle solute distribution function $g_{aa}$.

We emphasize that while the mean field approximation used in the theory curtails the solute fluctuations, it does not affect the solvent fluctuations. All basic features of the solvent density-density correlation function, $\chi_{bb}$, remain unaffected. Thus, when the system approaches to the solvent critical point amplitudes of the solvent density fluctuations and length over which local fluctuations are correlated will grow as they do in the bulk solvent. These critical fluctuations will therefore scale the solvent mediated potential, $v$, on length scale of the bulk correlation length $\xi$ which is known to diverge with critical exponent $\nu$. As a result, near the solvent critical point, the effective potential, $U_{aa}^{eff}$, between solute particles will develop features of what is known as the critical Casmir interaction \cite{maciolek2018collective}.

The theory is applied in Sec.\ref{Results} to calculate properties of the AO model which describes colloidal hard-spheres dissolved in a solvent of non interacting point particles.
The results plotted in Figs.1-9 and given in the table give values of changes that have taken place due to many-body effect on the effective attraction between colloidal particles and on the solute-solute and solute-solvent correlation functions. The theory can be extended straightforwardly to include features of real systems such as multi-component solvent and non-spherical molecules. It is self-contained in the sense that all quantities appearing in it are calculated from the microscopic interaction between particles of the full systems and can be used to study a variety of solute-solvent systems including the colloidal suspension with near-critical solvent.

\begin{table}[H]
\centering
\begin{tabular}{ | m{4em} | m{4em} | m{4em} | m{4em}| m{4em} | m{4em} |  m{4em} | } 
\hline
\multicolumn{4}{|c|}{$\eta_{b}=0.2$} &  \multicolumn{3}{c|}{$\eta_{b}=0.3$} \\ 
\hline
\ q & $\ v_{AO}$  & \multicolumn{2}{c|}{$\Delta v $} &  $\ v_{AO}$  & \multicolumn{2}{c|}{$\Delta v$} \\ 
\cline{3-4}
\cline{6-7}

	&  & $\eta_{a}=0.1$ &  $\eta_{a}=0.2$ & &  $\eta_{a}=0.1$ & $\eta_{a}=0.2$ \\  
\hline
0.25 & -1.40 & 0.12 & 0.24 & -2.11 & 0.25 & 0.43  \\ 
0.50 & -0.80 & 0.18 & 0.31 & -1.21 & 0.30 & 0.49  \\
0.75 & -0.60 & 0.23 & 0.36 & -0.90 & 0.36 & 0.54  \\
\hline
\end{tabular}
\vspace{1cm}
\caption{Values of $v(\frac{R}{\sigma_{a}}=1,\eta_{a},q)=v_{AO}$ and of the difference $\Delta v(\frac{R}{\sigma_{a}}=1,\eta_{a},q)=v(\frac{R}{\sigma_{a}}=1,\eta_{a},q)-v(\frac{R}{\sigma_{a}}=1,\eta_{a}=0,q)$ at $\eta_{b}=0.2$ and $0.3$ are given for $q=0.25, 0.5$ and $0.75$ and for $\eta_{a}=0.1$ and $0.2$. The difference $\Delta v$ measures the many-body effect on the effective attraction (depletion potential) between colloidal particles at contact. }
\label{table:1}
\end{table}

\section*{Conflicts of interest}
There are no conflicts to declare.

\begin{acknowledgments}
One of us (M.Y.) thanks the University Grants Commission, New Delhi, India, for award of research fellowship. We thank Referees for their comments and suggestions.
\end{acknowledgments}

\bibliographystyle{rsc}
\bibliography{Draft}

\end{document}